\documentstyle[seceq,mbf,wrapft]{ptptex}

\def\br{{\mbf r}}

\def\bfrho{{\mbf \rho}}
\def\bflambda{{\mbf \lambda}}

\def\bX{{\mbf X}}

\def\CO{{\cal O}}

\def\SM3{\Sigma N (3/2)}
\def\SN1{\Sigma N (1/2)}

\def\TS1{\hbox{}^3S_1}
\def\TD1{\hbox{}^3D_1}

\def\CO{{\cal O}}

\def\eq#1{Eq.~(\ref{#1})}

\def\CVwe{{\cal V}(\omega, \varepsilon)}

\def\VRGM{V^{\rm RGM}(\varepsilon)}
\def\VRGMA{V^{\rm RGM}_\alpha(\varepsilon_\alpha)}
\def\VRGMB{V^{\rm RGM}_\beta(\varepsilon_\beta)}
\def\VRGMC{V^{\rm RGM}_\gamma(\varepsilon_\gamma)}
\def\va{v_\alpha(\varepsilon_\alpha)}
\def\vb{v_\beta(\varepsilon_\beta)}
\def\vc{v_\gamma(\varepsilon_\gamma)}
\def\CVA{{\cal V}^{(3)}_\alpha(E, \varepsilon_\alpha)}
\def\CVB{{\cal V}^{(3)}_\beta(E, \varepsilon_\beta)}
\def\CVC{{\cal V}^{(3)}_\gamma(E, \varepsilon_\gamma)}
\def\TTE{\widetilde{T}^{(3)}_\alpha(E, \varepsilon_\alpha)}

\preprintnumber{
KUNS-1749 \\
nucl-th/0112070 \\ December 2001}
\title{Three-Cluster Equation Using Two-Cluster RGM Kernel}

\author{
Yoshikazu {\sc Fujiwara}, Hidekatsu {\sc Nemura}$^{*}$,
Yasuyuki {\sc Suzuki}$^{**}$ \\
Kazuya {\sc Miyagawa}$^{***}$ and Michio {\sc Kohno},$^{****}$
}

\inst{
Department of Physics, Kyoto University,
Kyoto 606-8502, Japan \\
$\hbox{}^{*}$Institute of Particle and Nuclear Physics, KEK,
Ibaragi 305-0801, Japan \\
$\hbox{}^{**}$Department of Physics, Niigata University,
Niigata 950-2181, Japan \\
$\hbox{}^{***}$Department of Applied Physics, Okayama Science University,
Okayama 700-0005, Japan \\
$\hbox{}^{****}$Physics Division, Kyushu Dental College,
Kitakyushu 803-8580, Japan
}
\recdate{
\today
}
\abst{
We propose a new type of three-cluster equation
which uses two-cluster resonating-group-method (RGM) kernels.
In this equation, the orthogonality
of the total wave function to two-cluster Pauli-forbidden states
is essential to eliminate redundant components admixed
in the three-cluster systems. The explicit energy-dependence
inherent in the exchange RGM kernel is self-consistently determined.
For bound-state problems, this equation is straightforwardly
transformed to the Faddeev equation which uses a modified
singularity-free $T$-matrix constructed from the two-cluster RGM kernel.
The approximation of the present three-cluster formalism can be examined
with more complete calculation using the three-cluster RGM.
As a simple example, we discuss three di-neutron ($3d^\prime$)
and $3\alpha$ systems in the harmonic-oscillator variational calculation.
The result of the Faddeev calculation is also presented
for the $3d^\prime$ system.
}

\begin{document}

\maketitle

\section{Introduction}

All the present-day quark-model descriptions
of the nucleon-nucleon ($NN$) and hyperon-nucleon ($YN$) interactions
incorporate important roles of the quark-gluon degrees of freedom
in the the short-range region and the meson-exchange processes
dominated in the medium- and long-range parts of the interaction. \cite{OY84}
For example, we have introduced one-gluon exchange Fermi-Breit interaction
and effective meson-exchange potentials acting between quarks,
and have achieved very accurate descriptions
of the $NN$ and $YN$ interactions
with limited number of parameters. \cite{FU96a,FU96b,FU01b}
We hope that the derived interaction in these models
can be used for a realistic calculation
of few-baryon systems like the hypertriton
and various types of baryonic matter.
This program, however, involves a non-trivial problem
of how to extract the effective two-baryon interaction
from the microscopic quark-exchange kernel.
The basic baryon-baryon interaction is formulated as a composite-particle
interaction in the framework of the resonating-group method (RGM).
If we rewrite the RGM equation in the form
of the Schr{\" o}dinger-type equation,
the interaction term becomes non-local and energy dependent.
Furthermore, the RGM equation sometimes involves
redundant components due to the effect of
the antisymmetrization, which is related to the existence
of the Pauli-forbidden states. 
In such a case, the full off-shell $T$-matrix is not well
defined in the standard procedure which usually assumes
simple energy-independent local potentials. \cite{GRGM}
Since these features are related to the characteristic
description of the short-range part in the quark model,
it would be desirable if one can use the quark-exchange kernel directly
in the application to many-baryon systems.

In this paper, we propose a new type of three-cluster equation
which uses two-cluster RGM kernel for the inter-cluster interaction.
We assume, for simplicity, three identical clusters
having only one Pauli-forbidden state for the inter-cluster relative motion,
but the extension to general systems is rather straightforward.
We first consider a two-cluster RGM equation and the structure
of $T$-matrix constructed from the two-cluster RGM kernel (RGM $T$-matrix).
Next, we formulate a three-cluster equation which employs the two-cluster
RGM kernel and a projection operator on the pairwise Pauli-allowed space.
This equation is converted to the Faddeev equation which uses
a non-singular part of the RGM $T$-matrix.
Finally, we show some examples of the present formulation
with respect to the $0^+$ ground states of
three di-neutron ($3d^\prime$) and $3\alpha$ systems.
The calculation is performed in the variational method,
using the translationally-invariant harmonic oscillator basis.
For the $3d^\prime$ system, the result of the Faddeev calculation
is also presented.
Detailed comparison is made with respect to the more desirable
three-cluster RGM calculation, and to some other approximations
like ``renormalized RGM'' and the well-known
orthogonality condition model (OCM). \cite{SA68,SA77,HO74}

\section{$T$-matrix of the two-cluster RGM kernel}

We use the same notation as used in Ref.\,\citen{GRGM} and
write a two-cluster RGM equation as
\begin{eqnarray}
\left[\,\varepsilon-H_0-V^{\rm RGM}(\varepsilon)\,\right] \chi=0\ ,
\label{form1}
\end{eqnarray}
where $\varepsilon$ is the relative energy, $\varepsilon=E-E^{\rm int}$,
between the two clusters,
$H_0$ is the relative kinetic-energy operator, and
\begin{eqnarray}
\VRGM=V_{\rm D}+G+\varepsilon K\ ,
\label{form2}
\end{eqnarray}
is the RGM kernel composed of the direct potential $V_{\rm D}$, the sum
of the exchange kinetic-energy and interaction kernels,
$G=G^{\rm K}+G^{\rm V}$,
and the exchange normalization kernel $K$. We assume that there exists
only one Pauli-forbidden state $|u\rangle$, which satisfies the
eigen-value equation $K |u\rangle=\gamma |u\rangle$ with
the eigen-value $\gamma=1$.
The projection operator on the Pauli-allowed space for the relative
motion is denoted by $\Lambda=1-|u\rangle \langle u|$.
Using the basic property of the Pauli-forbidden state $|u\rangle$,
$(H_0+V_{\rm D}+G)|u\rangle=\langle u|(H_0+V_{\rm D}+G)=0$,
we can separate $\VRGM$ into two distinct parts:
\begin{eqnarray}
\VRGM=V(\varepsilon)+v(\varepsilon)\ ,
\label{form3}
\end{eqnarray}
where
\begin{eqnarray}
& & V(\varepsilon)=(\varepsilon-H_0)-\Lambda (\varepsilon-H_0) \Lambda
=\varepsilon |u\rangle \langle u| +\Lambda H_0 \Lambda-H_0\ ,\nonumber \\
& & v(\varepsilon)=\Lambda \VRGM \Lambda
=\Lambda \left( V_{\rm D}+G+\varepsilon K\right) \Lambda\ .
\label{form4}
\end{eqnarray}
Note that $\Lambda V(\varepsilon) \Lambda=0$
and $\Lambda v(\varepsilon) \Lambda=v(\varepsilon)$; namely,
$V(\varepsilon)$ may be considered as an operator acting in the
Pauli-forbidden space, while $v(\varepsilon)$ an operator
acting in the Pauli-allowed space.
Using these properties, we can express \eq{form1} as
\begin{eqnarray}
\Lambda \left[\,\varepsilon-H_0-v(\varepsilon)\,\right] \Lambda \chi=0\ .
\label{form5}
\end{eqnarray}
The separation of $\VRGM$ in \eq{form3} enables us to deal with the
energy dependence of the exchange RGM kernel in the Pauli-forbidden
space and that in the allowed space, separately.
Let us generalize \eq{form3} to
\begin{eqnarray}
\CVwe=V(\omega)+v(\varepsilon)\ ,
\label{form6}
\end{eqnarray}
which we use in the following three-cluster formulation.
We will see that the energy dependence involved $V(\omega)$ can be
eliminated by the orthogonality condition
to the Pauli-forbidden state.
  
Since the direct application of the $T$-matrix formalism to \eq{form1}
leads to a singular off-shell $T$-matrix, \cite{GRGM}
we first consider the subsidiary equation
\begin{eqnarray}
\left[\,\omega -H_0-\VRGM\,\right] \chi=0\ ,
\label{form7}
\end{eqnarray}
with $\omega \neq \varepsilon$,
and extract a singularity-free
off-shell $T$-matrix starting from the standard $T$-matrix formulation
for the ``potential'' $\VRGM$.
A formal solution of the $T$-matrix equation
\begin{eqnarray}
T(\omega,\varepsilon)=\VRGM+\VRGM G^{(+)}_0(\omega)
T(\omega,\varepsilon)
\label{form8}
\end{eqnarray}
with $G^{(+)}_0(\omega)=1/(\omega-H_0+i0)$ is given by
\begin{eqnarray}
T(\omega,\varepsilon) & = & \widetilde{T}(\omega, \varepsilon)
+(\omega-H_0)|u\rangle {1 \over \omega-\varepsilon}
\langle u|(\omega-H_0) \ ,\nonumber \\
\widetilde{T}(\omega, \varepsilon) & = & T_v(\omega, \varepsilon)
-\left(1+T_v(\omega, \varepsilon) G^{(+)}_0(\omega)\right)|u \rangle
{1 \over \langle u|G^{(+)}_v(\omega, \varepsilon)|u \rangle} \nonumber \\
& & \times \langle u| \left(1+G^{(+)}_0(\omega)
T_v(\omega, \varepsilon)\right) \ ,
\label{form9}
\end{eqnarray}
where $T_v(\omega, \varepsilon)$ is defined by
\begin{eqnarray}
T_v(\omega, \varepsilon)=v(\varepsilon)+v(\varepsilon) G^{(+)}_0(\omega)
T_v(\omega, \varepsilon) \ .
\label{form10}
\end{eqnarray}
This result is obtained through the expression for the full Green function
given by 
\begin{eqnarray}
G^{(+)}(\omega,\varepsilon)={1 \over \omega-H_0-\VRGM+i0}
=G^{(+)}_{\Lambda}(\omega, \varepsilon)
+|u\rangle {1 \over \omega-\varepsilon}
\langle u|\ ,\qquad
\label{form11}
\end{eqnarray}
where
\begin{eqnarray}
G^{(+)}_{\Lambda}(\omega, \varepsilon)
& = & G^{(+)}_v(\omega, \varepsilon)-G^{(+)}_v(\omega, \varepsilon)|u\rangle
{1 \over \langle u|G^{(+)}_v(\omega, \varepsilon)|u \rangle} \nonumber \\
& & \times \langle u|G^{(+)}_v(\omega, \varepsilon)\ ,
\label{form12}
\end{eqnarray}
and $G^{(+)}_v(\omega, \varepsilon)
=1/(\omega-H_0-v(\varepsilon)+i0)$ is the solution of
\begin{eqnarray}
G^{(+)}_v(\omega, \varepsilon)=G^{(+)}_0(\omega)+G^{(+)}_0(\omega)
\,v(\varepsilon) G^{(+)}_v(\omega, \varepsilon)\ .
\label{form13}
\end{eqnarray}
In fact, the simple relationship
\begin{eqnarray}
V(\varepsilon)=V(\omega)-(\omega-\varepsilon)|u\rangle \langle u|
\label{form13-1}
\end{eqnarray}
yields
\begin{eqnarray}
\omega-H_0-\VRGM=\Lambda \left(\omega-H_0-v(\varepsilon)\right)\Lambda
+(\omega-\varepsilon)|u\rangle \langle u|\ .
\label{form13-2}
\end{eqnarray}
Then, if one uses the property
\begin{eqnarray}
& & \Lambda \left[\,\omega-H_0-v(\varepsilon)\,\right]\Lambda
\cdot G^{(+)}_{\Lambda}(\omega, \varepsilon) \nonumber \\
& & =G^{(+)}_{\Lambda}(\omega, \varepsilon) \cdot
\Lambda \left[\,\omega-H_0-v(\varepsilon)\,\right]\Lambda=\Lambda\ ,
\label{form13-3}
\end{eqnarray}
it is easily found that
\begin{eqnarray}
& & \left[\,\omega-H_0-\VRGM\,\right]
\ G^{(+)}(\omega, \varepsilon) \nonumber \\
& & =\left\{\,\Lambda \left[\,\omega-H_0-v(\varepsilon)\,\right]\Lambda
+(\omega-\varepsilon)|u\rangle \langle u|\,\right\}
\left\{\,G^{(+)}_{\Lambda}(\omega, \varepsilon)
+|u\rangle {1 \over \omega-\varepsilon}
\langle u|\,\right\} \nonumber \\
& & =\Lambda+|u\rangle \langle u|=1\ .
\label{form13-4}
\end{eqnarray}
The expression of $T(\omega,\varepsilon)$ in \eq{form9} is most easily
obtained from
\begin{eqnarray}
G^{(+)}(\omega, \varepsilon)=G^{(+)}_0(\omega)
+G^{(+)}_0(\omega)\,T(\omega,\varepsilon)\,G^{(+)}_0(\omega)\,
\label{form13-5}
\end{eqnarray}
or
\begin{eqnarray}
T(\omega,\varepsilon)=\left(\omega-H_0\right)
\,G^{(+)}(\omega, \varepsilon)\,\left(\omega-H_0\right)
-\left(\omega-H_0\right)\ .
\label{form13-6}
\end{eqnarray}

The basic relationship which will be used in the following is 
\begin{eqnarray}
G^{(+)}_0(\omega) T(\omega,\varepsilon)
& = & G^{(+)}(\omega, \varepsilon ) \VRGM \nonumber \\
& = & G^{(+)}_\Lambda(\omega, \varepsilon) \VRGM
+|u\rangle {1 \over \omega-\varepsilon}
\langle u|\VRGM \nonumber \\
& = & G^{(+)}_\Lambda(\omega, \varepsilon) \CVwe
-|u\rangle \langle u|
+|u\rangle {1 \over \omega-\varepsilon}
\langle u|(\omega-H_0) \nonumber \\
& = &G^{(+)}_0(\omega) \widetilde{T}(\omega, \varepsilon)
+|u\rangle {1 \over \omega-\varepsilon}
\langle u|(\omega -H_0)\ ,
\label{form14}
\end{eqnarray}
where $\widetilde{T}(\omega, \varepsilon)$ satisfies
\begin{eqnarray}
G^{(+)}_0(\omega) \widetilde{T}(\omega, \varepsilon)
=G^{(+)}_\Lambda(\omega, \varepsilon) \CVwe
-|u\rangle \langle u| \ .
\label{form15}
\end{eqnarray}
These can be shown by using Eqs.\,(\ref{form11}) and
(\ref{form13-1}).
The full $T$-matrix, $T(\omega, \varepsilon)$,
in \eq{form9} is singular at $\varepsilon=\omega$,
while $\widetilde{T}(\omega, \varepsilon)$ does not
have such singularity.
For $\varepsilon \neq \omega$, $T(\omega, \varepsilon)$ satisfies
the relationship
\begin{eqnarray}
\langle u|G^{(+)}_0(\omega) T(\omega,\varepsilon)|\phi \rangle
=\langle \phi | T(\omega,\varepsilon) G^{(+)}_0(\omega)
|u \rangle=0
\label{form16}
\end{eqnarray}
for the plane wave solution $|\phi \rangle$ with the energy $\varepsilon$,
i.e., $(\varepsilon-H_0)|\phi \rangle=0$.
This relationship is a direct result of more general relationship
\begin{eqnarray}
\langle u| \left[\,1+G^{(+)}_0(\omega) \widetilde{T}(\omega, \varepsilon)
\,\right]=\left[\, 1+\widetilde{T}(\omega, \varepsilon)
G^{(+)}_0(\omega)\,\right] |u \rangle=0 \ ,
\label{form17}
\end{eqnarray}
which is simply seen from \eq{form15}.

We note that $\widetilde{T}(\omega, \varepsilon)$ satisfies
\begin{eqnarray}
& & \widetilde{T}(\omega, \varepsilon)=\CVwe
-|u\rangle \langle u |(\omega-H_0)+\CVwe G^{(+)}_0(\omega)
\widetilde{T}(\omega, \varepsilon)\ , \nonumber \\
& & \widetilde{T}(\omega, \varepsilon)=\CVwe
-(\omega-H_0)|u\rangle \langle u |+\widetilde{T}(\omega, \varepsilon)
G^{(+)}_0(\omega) \CVwe \ .
\label{form18}
\end{eqnarray}
However, these asymmetric forms of the $T$-matrix equations do not
determine the solution $\widetilde{T}(\omega, \varepsilon)$ uniquely,
since the resolvent kernel $\left[\,1-\CVwe G^{(+)}_0(\omega)\,\right]^{-1}$
has a singularity related to the existence
of the trivial solution $|u\rangle$:
\begin{eqnarray}
\langle u|\left[\,1-\CVwe G^{(+)}_0(\omega)\,\right]=0\ .
\label{form19}
\end{eqnarray}
The driving term, $\CVwe-|u\rangle \langle u |(\omega-H_0)$, etc., guarantees
the existence of the solution, except for an arbitrary admixture
of $(\omega-H_0)|u\rangle$ component. In order to eliminate this ambiguity
and to make $\widetilde{T}(\omega, \varepsilon)$ symmetric,
one has to impose some orthogonality conditions, which will be
discussed in a separate paper.

\section{Three-cluster equation}

Let us consider a system composed of three identical spinless particles,
interacting via the two-cluster RGM kernel $\VRGM$.
The energy dependence involved in $\VRGM$ should be
treated properly by calculating the expectation value
of the two-cluster subsystem, at least for $v(\varepsilon)$.
On the other hand, the energy dependence involved in $V(\varepsilon)$ is
of kinematical origin, and could be modified so as to be best
suited to the three-cluster equation.
The three-body equation we propose is expressed as
\begin{eqnarray}
P \left[\,E-H_0-\VRGMA-\VRGMB-\VRGMC\,\right] P \Psi=0\ ,
\label{three1}
\end{eqnarray}
where $H_0$ is the free three-body kinetic-energy operator
and $\VRGMA$ stands for the RGM kernel \eq{form2} for
the $\alpha$-pair, etc.
The two-cluster relative energy $\varepsilon_\alpha$ in the
three-cluster system is self-consistently determined through
\begin{eqnarray}
\varepsilon_\alpha=\langle P \Psi|
\,h_\alpha+\VRGMA\,|P \Psi \rangle\ ,
\label{three1-2}
\end{eqnarray}
using the normalized three-cluster wave
function $P \Psi$ with $\langle P \Psi| P \Psi \rangle=1$.
Here $h_\alpha$ is the free kinetic-energy operator for the $\alpha$-pair.
Also, $P$ is the projection operator on the [3] symmetric Pauli-allowed
space as defined below. \cite{HO74}
We solve the eigen-value problem
\begin{eqnarray}
\sum_\alpha|u_\alpha \rangle \langle u_\alpha | \Psi_\lambda \rangle
=\lambda~| \Psi_\lambda \rangle
\label{three2}
\end{eqnarray}
in the [3] symmetric model space, $|\Psi_\lambda \rangle \in [3]$,
and define $P$ as a projection on the space spanned
by eigen-vectors with the eigen-value $\lambda=0$:
\begin{eqnarray}
P=\sum_{\lambda=0}|\Psi_\lambda \rangle \langle \Psi_\lambda |\ .
\label{three3}
\end{eqnarray}
It is easy to prove that $P$ has the following properties:
\begin{eqnarray}
&({\rm i}) &~\Lambda_\alpha P=P \Lambda_\alpha=P \quad \hbox{for} \quad
\hbox{}^\forall \alpha\ ,\nonumber \\
&({\rm ii}) &~\hbox{when}~\Psi \in [3],
~\hbox{}^\forall~\langle u_\alpha|\Psi \rangle=0
\longleftrightarrow  P \Psi=\Psi\ ,\nonumber \\
&({\rm iii}) &~\hbox{when}~\Psi \in [3],~P \Psi=0 
\longleftrightarrow \hbox{}^\exists |f\rangle\ ,\nonumber \\
& & ~\hbox{such~that}
~\Psi=|u_\alpha\rangle |f_\alpha\rangle+|u_\beta\rangle |f_\beta\rangle
+|u_\gamma\rangle |f_\gamma\rangle\ .
\label{three4}
\end{eqnarray}
Note that all these relations are considered in the [3] symmetric
model space, and $P$ and $Q \equiv 1-P$ are both [3] symmetric
three-body operators.
Using the property (i), we can simplify \eq{three1} as
\begin{eqnarray}
P \left[\,E-H_0-\va-\vb-\vc\,\right] P \Psi=0\ .
\label{three5}
\end{eqnarray}

In order to derive the Faddeev equation corresponding to \eq{three1},
it is convenient to rewrite \eq{three1} or (\ref{three5}) as
\begin{eqnarray}
P \left[\,E-H_0-\CVA-\CVB-\CVC\,\right] P \Psi=0\ ,
\label{three6}
\end{eqnarray}
where $\CVA$ is the three-body operator defined
by $\CVwe$ in \eq{form6} through
\begin{eqnarray}
\CVA & = & {\cal V}_\alpha (E-h_{\bar \alpha}, \varepsilon_\alpha)
\nonumber \\
& = & (E-H_0)-\Lambda_\alpha (E-H_0) \Lambda_\alpha
+v_\alpha(\varepsilon_\alpha)\ .
\label{three7}
\end{eqnarray}
Here $h_{\bar \alpha}$ is the kinetic-energy operator between
the $\alpha$-pair and the third particle.
The last equation of \eq{three7} is derived
by using $h_\alpha+h_{\bar \alpha}=H_0$.
The validity of \eq{three6} is easily seen from, for example,
$P\CVB P=P \Lambda_\beta \CVB \Lambda_\beta P
=P v_\beta(\varepsilon_\beta) P$,
which uses the property (i) of \eq{three4}.
The expression behind the first $P$ in the left-hand side of \eq{three6}
is symmetric with respect to the exchange of the three particles.
\footnote{The two-cluster relative energies, $\varepsilon_\alpha$, 
$\varepsilon_\beta$ and $\varepsilon_\gamma$, are actually all equal,
since we are dealing with the three identical particles.}
Thus, by applying the property (iii) of \eq{three4}, we find
\begin{eqnarray}
& & \left[\,E-H_0-\CVA-\CVB-\CVC\,\right] P \Psi \nonumber \\
& & =|u_\alpha\rangle |f_\alpha\rangle+|u_\beta\rangle |f_\beta\rangle
+|u_\gamma\rangle |f_\gamma\rangle\ ,
\label{three8}
\end{eqnarray}
where $|f\rangle$ is an unknown function, and $|f_\beta\rangle$
and $|f_\gamma\rangle$ are simply obtained from $|f_\alpha\rangle$ by
the cyclic permutations.
Here we invoke the standard ansatz to set
\begin{eqnarray}
P \Psi=\psi_\alpha+\psi_\beta+\psi_\gamma\ ,
\label{three9}
\end{eqnarray}
and define $\psi_\alpha$ as the solution of
\begin{eqnarray}
\left(\,E-H_0\,\right) \psi_\alpha
=\CVA P \Psi+|u_\alpha\rangle |f_\alpha\rangle\ .
\label{three10}
\end{eqnarray}
This equation can be written as
\begin{eqnarray}
\left[\,E-H_0-\CVA\,\right] \psi_\alpha
=\CVA (\psi_\beta+\psi_\gamma)+|u_\alpha\rangle |f_\alpha\rangle\ ,
\label{three11}
\end{eqnarray}
or by using \eq{three7} as
\begin{eqnarray}
\Lambda_\alpha \left[\,E-H_0-\va\,\right] \Lambda_\alpha \psi_\alpha
=\CVA (\psi_\beta+\psi_\gamma)+|u_\alpha\rangle |f_\alpha\rangle\ .
\label{three12}
\end{eqnarray}
The unknown function $|f_\alpha\rangle$ is determined
if we multiply this equation by $\langle u_\alpha|$ from the left
and use $\langle u_\alpha|\CVA=\langle u_\alpha|(E-H_0)$:
\begin{eqnarray}
|f_\alpha\rangle=-\langle u_\alpha|\,E-H_0\,| \psi_\beta+\psi_\gamma
\rangle\ .
\label{three13}
\end{eqnarray}
Thus we obtain
\begin{eqnarray}
\Lambda_\alpha \left[\,E-H_0-\va\,\right] \Lambda_\alpha \psi_\alpha
& = & \CVA (\psi_\beta+\psi_\gamma) \nonumber \\
& & -|u_\alpha\rangle \langle u_\alpha|\,E-H_0\,| \psi_\beta+\psi_\gamma
\rangle\ .
\label{three14}
\end{eqnarray}
By employing the two-cluster relation \eq{form15} in the three-cluster
model space,
\begin{eqnarray}
G^{(+)}_{\Lambda_\alpha}(E, \varepsilon_\alpha) \CVA
=G^{(+)}_0(E) \TTE + |u_\alpha\rangle \langle u_\alpha| \ ,
\label{three15}
\end{eqnarray}
and the relationship,
\begin{eqnarray}
& & G^{(+)}_{\Lambda_\alpha}(E, \varepsilon_\alpha)
\,\Lambda_\alpha \left[\,E-H_0-\va\,\right] \Lambda_\alpha
\nonumber \\
& & =\Lambda_\alpha \left[\,E-H_0-\va\,\right] \Lambda_\alpha
\,G^{(+)}_{\Lambda_\alpha}(E, \varepsilon_\alpha)=\Lambda_\alpha\ ,
\label{three16}
\end{eqnarray}
\eq{three14} yields
\begin{eqnarray}
\Lambda_\alpha \psi_\alpha
=G^{(+)}_0(E) \TTE (\psi_\beta+\psi_\gamma)
+|u_\alpha\rangle \langle u_\alpha| \psi_\beta+\psi_\gamma \rangle\ .\qquad
\label{three17}
\end{eqnarray}
Since $\langle u_\alpha| \psi_\beta+\psi_\gamma\rangle=
-\langle u_\alpha| \psi_\alpha\rangle$ from \eq{three9}, we finally
obtain
\begin{eqnarray}
\psi_\alpha=G^{(+)}_0(E) \TTE (\psi_\beta+\psi_\gamma)\ .
\label{three18}
\end{eqnarray}
Note that $\TTE$ is essentially the non-singular part of the
two-cluster RGM $T$-matrix \eq{form9}:
\begin{eqnarray}
\TTE=\widetilde{T}_\alpha(E-h_{\bar \alpha}, \varepsilon_\alpha)\ ,
\label{three19}
\end{eqnarray}
and that the solution of \eq{three18} automatically satisfies
$\langle u_\alpha| \psi_\alpha+\psi_\beta+\psi_\gamma\rangle=0$ due
to \eq{form17}.
Since $\psi_\alpha+\psi_\beta+\psi_\gamma \in [3]$,
the property (ii) of \eq{three4} yields $\Psi=P \Psi$ if
we set $\Psi=\psi_\alpha+\psi_\beta+\psi_\gamma$.
We can also start from \eq{three18} and derive \eq{three1}
by using the properties (i) and (ii) of \eq{three4},
thus establish the equivalence between \eq{three1} and \eq{three18}.

\section{Three di-neutron system}

As a simplest non-trivial example, we first consider
three di-neutron ($d^\prime$) system, where the internal wave function
of the $d^\prime$ is assumed to be $(0s)$ harmonic oscillator (h.o.)
wave function. The normalization kernel $K$ for the $2d^\prime$ system
is given by $K=\Lambda=1-|u\rangle \langle u|$ and the $\Lambda
(\varepsilon K) \Lambda$ term in $v(\varepsilon)$ disappears.
Here $|u\rangle$ is the $(0s)$ h.o. wave function
given by $u(\br)=\langle \br|u\rangle
=(2\nu/\pi)^{3/4} e^{-\nu \br^2}$.
We assume a very simple two-nucleon interaction of the Serber type
\begin{eqnarray}
v_{ij}=-v_0~e^{-\kappa r^2}~{1 \over 2}\left(1+P_r\right) \ ,
\label{deut1}
\end{eqnarray}
according to Ref.\,\citen{SA73}. This paper deals with a schematic model
of the almost forbidden state with $v_0=90$ MeV,
but this strength is too weak to give a bound state
for the $3d^\prime$ system. We use the following parameter set
in the present calculation: $\nu=0.12~\hbox{fm}^{-2}$,
$\kappa=0.46~\hbox{fm}^{-2}$ and $v_0=153~\hbox{MeV}$.
With this value of $v_0$, the $2d^\prime$ system is 
slightly bound.\footnote{The $S$-wave phase shift
for the $2d^\prime$ scattering shows that the $2d^\prime$ system
gets bound between $v_0=151$ MeV and 152 MeV.}

In order to solve the three-cluster equation (\ref{three1}),
we first prepare [3] symmetric translationally-invariant h.o. basis
according to the Moshinsky's method \cite{MO96}.\footnote{This process
is most easily formulated using the theory
of Double Gel'fand polynomials. \cite{KI83}.}
The [3] symmetric Pauli-allowed states,
which we denote by $\varphi^{[3](\lambda \mu)}_{a, n}(\bfrho, \bflambda)$,
are explicitly constructed
by the diagonalization procedure in \eq{three2}.
Here, $\bfrho=(\bX_1-\bX_2)/\sqrt{2}$ and $\bflambda
=(\bX_1+\bX_2-2\bX_3)/\sqrt{6}$ are the Jacobi coordinates
for the center-of-mass coordinates $\bX_i$ ($i=1$ - 3) of
the three $d^\prime$ clusters.
These eigen-states are specified by the $SU_3$ quantum
number $(\lambda \mu)$ and a set of the other quantum numbers $n$,
which includes the total h.o. quanta $N$.
We then perform the variational calculation using these basis states.
Namely, we first expand $P \Psi$ as
\begin{eqnarray}
P \Psi=\sum_{(\lambda, \mu), n} c^{(\lambda, \mu)}_n
\varphi^{[3](\lambda \mu)}_{a, n}(\bfrho, \bflambda)\ .
\label{deut2}
\end{eqnarray}
In the following, we use a simplified notation $n$ to represent
the set of $(\lambda \mu)$ and $n$ (and also the possible $K$ quantum
number if the total orbital angular momentum $L \neq 0$).
Since $\Psi$ is [3] symmetric, the three interaction
terms in \eq{three1} give the
same contribution. This leads to the eigen-value equation
\begin{eqnarray}
& & \sum_{n^\prime}\left(E~\delta_{nn^\prime}-H_{nn^\prime}\right)
c_{n^\prime}=0\ , \nonumber \\
& & H_{nn^\prime}=(H_0)_{nn^\prime}
+3 \left[\,(V_{\rm D})_{nn^\prime}+G_{nn^\prime}+\varepsilon K_{nn^\prime}
\,\right]\ .
\label{deut3}
\end{eqnarray}
Here $K_{nn^\prime}$ term in the $3d^\prime$ problem
is trivially zero since the [3] symmetric allowed basis does not
contain the $(0s)$ component from the very beginning.
This implies that our $d^\prime d^\prime$ interaction
is energy independent and the self-consistency
for $\varepsilon$ is not necessary.
On the other hand, the $3\alpha$ case which will be discussed in the next
section requires to determine $\varepsilon$ through
\begin{eqnarray}
\varepsilon={\sum_{n,n^\prime}\left[h_{nn^\prime}
+(V_{\rm D})_{nn^\prime}+G_{nn^\prime}\right] c_n c_{n^\prime} \over
1-\sum_{n,n^\prime}K_{nn^\prime} c_n c_{n^\prime}}\ .
\label{deut4}
\end{eqnarray}
The two-body matrix elements in the three-body
space, $\CO_{nn^\prime}=\langle \varphi^{[3]}_n |\CO|
\varphi^{[3]}_{n^\prime}\rangle$,
can be calculated by using the power series expansion
of the complex GCM kernel and $SU_3$ Clebsch-Gordan (C-G) coefficients
of the type $\langle (N_1 0) \ell (N_2 0) \ell || (\lambda \mu)
0 0 \rangle$. Fortunately, a concise expression is given by
Suzuki and Hecht \cite{SU83} for this particular type
of $SU_3$ C-G coefficients with $L=0$.

Table I shows the lowest eigen-values of \eq{deut3} with an increasing
number of total h.o. quanta $N$ included
in the calculation. The number of basis states
rapidly increases, as the maximum $N$ becomes larger.
The listing is terminated
when the number of basis states $n_{\rm Max}$ is over 1000,
which is reached around $N\sim 60$. The convergence
of the $3d^\prime$ system is rather slow, since the bound-state energy
is especially small in this particular system.
The best value obtained in the variational calculation
is $E_{3d^\prime}=-0.4323$ MeV,
using 2,927 basis states with $N \leq 88$.
We have also solved the Faddeev equation (\ref{three18}),
and obtained $E_{3d^\prime}=-0.4375$ MeV and $-0.4378$ MeV,
when the partial waves up to $\ell=4$ and 6 are included
in the calculation, respectively.
The final value $E_{3d^\prime}=-0.438$ MeV can also be compared
with $E^{\rm RGM}_{3d^\prime}=-1.188$ MeV,
which is obtained by the stochastic variational method \cite{SU98}
for the $3d^\prime$ RGM.
Our result by the three-cluster equation gives 0.75 MeV less bound,
compared with the full microscopic $3d^\prime$ RGM calculation.  

\section{$3\alpha$ system}

In this system, the structure of the $2\alpha$ normalization kernel $K$ is
more involved. In the relative $S$-wave we have two Pauli-forbidden
states, $(0s)$ and $(1s)$, while in the $D$-wave only
one $(0d)$ h.o. state is forbidden.
The relative motion between the two $\alpha$ clusters
starts from $N=4$ h.o. quanta
The eigen-value $\gamma_N$ for $K$ is given
by $\gamma_N=2^{2-N}-3\delta_{N, 0}$, which is 1 ($N=0$ or 2),
1/4 ($N=4$), 1/16 ($N=6$), $\cdots$ .
The rather large value $\gamma_4=1/4$ makes the self-consistent
procedure through \eq{deut4} very important.
For the two-body effective interaction,
we use the Volkov No.\,2 force with $m=0.59$,
following the $3\alpha$ RGM calculation
by Fukushima and Kamimura \cite{FU78}.
The h.o. constant for the $\alpha$ cluster is $\nu=0.275~\hbox{fm}^{-2}$,
which gives the $\alpha$ cluster internal energy $E_\alpha=-27.3$ MeV
for the $(0s)^4$ configuration, if the Coulomb interaction is included.
(Cf. $E^{\rm exp't}_\alpha=-28.3$ MeV.)
We have carried out the $2\alpha$ RGM calculation by using this
parameter set, and found that the present $2\alpha$ system is
bound with the binding energy $E_{2\alpha}=-0.245$ MeV.
(Cf. $E^{\rm exp't}_{2\alpha}=92$ keV.)

Table I lists the convergence of the lowest $3\alpha$ eigen-values
with respect to the maximum total h.o. quanta $N$.
We find that the final values of $E_{3\alpha}$ and $\varepsilon$
are $E_{3\alpha}=-5.97$ MeV and $\varepsilon=9.50$ MeV.
If we compare this with $E^{\rm RGM}_{3\alpha}=-7.5$ MeV
(Cf. $E^{\rm exp't}_{3\alpha}=-7.3$ MeV) by the $3\alpha$ RGM
calculation \cite{FU78}, we find that our result is 1.5 MeV less bound.
The amplitude of the lowest shell-model component with the $SU_3$ (04)
representation is $c_{(04)}=0.790$.
We think that the underbinding compared to the three-cluster
RGM calculation is reasonable, since our three-cluster equation
misses some attractive effect due to the genuine three-cluster
exchange kernel.
Oryu et al. carried out $3\alpha$ Faddeev calculation
using $2\alpha$ RGM kernel. \cite{OR94}
They obtained very large binding energy, $E_{3\alpha}=-28.2$ MeV
with the Coulomb force turned off.
Since the effect of the Coulomb force is at most 6 MeV,\footnote{In our
calculation, $E_{3\alpha}=-11.42$ MeV when the Coulomb interaction
is turned off, which implies that the effect of the Coulomb
interaction is 5.45 MeV.} this value is too deep.
This is because they did not treat the $\varepsilon K$ term
in the RGM kernel properly and the effect of $P$ in \eq{three1} is
not fully taken into account in their Faddeev formalism.
In order to see the importance of the $\varepsilon K$ term
in \eq{deut3}, it is useful to see the contribution of this term
in the lowest h.o. (04) configuration.
The decomposition of $E^{(04)}_{3\alpha}=12.634$ MeV with $N=8$ in
Table I is
\begin{eqnarray}
H & = & H_0+3\,(V_{\rm D}+G^{\rm K}+G^{\rm V}+{V_{\rm D}}^{\rm CL}
+G^{\rm CL}+\varepsilon K) \nonumber \\
  & = & 125.4+3(-36.54-15.68+6.54+2.58-0.78+25.12/4) \nonumber \\
  & = & 125.4-3 \times 37.6=12.6~\hbox{MeV}\ .
\label{alpha1}
\end{eqnarray}
This example shows very clearly that the self-consistent procedure 
for the energy-dependence of the RGM kernel in the allowed model
space is sometimes very important.

Since the present calculation employs the h.o. basis, it is very
easy to examine another approximation which eliminate
the explicit energy dependence involved in $v(\varepsilon)$.
This approximation is related to the proper normalization
of the two-cluster relative wave function $\chi$ in \eq{form1}
through $\psi=\sqrt{1-K}\chi$, and we call this approximation
the renormalized RGM. In this formulation, the interaction
generated from the RGM kernel is expressed as
\begin{eqnarray}
v=\left({1 \over \sqrt{1-K}}\right)^\prime\,(h_0+V_{\rm D}+G)
\,\left({1 \over \sqrt{1-K}}\right)^\prime - \Lambda h_0 \Lambda\ ,
\label{alpha2}
\end{eqnarray}
where the prime in $(1/\sqrt{1-K})^\prime$ implies the inversion
of $\sqrt{1-K}$ in the allowed model space.
(See Ref.\,\citen{SA77} and the discussion in Ref.\,\citen{GRGM}.)
The column $E^{\rm RN}_{3\alpha}$ in Table I shows the result
of this approximation. We find $E^{\rm RN}_{3\alpha}=-4.99$ MeV,
which is 0.98 MeV less bound in comparison with our result.
This may result from rather inflexible choice
of the $3\alpha$ Hamiltonian, caused by the lack
of the self-consistency. Table I also shows the result
by $3\alpha$ OCM ($E^{\rm OCM}_{3\alpha}$), whose procedure is
to use $v=\Lambda (V_{\rm D}+V^{\rm CL}_{\rm D}) \Lambda$.
We find $E^{\rm OCM}_{3\alpha}=-4.68$ MeV, which is
further 0.31 MeV less bound. In this case, $2\alpha$ OCM gives
a larger binding energy, $E^{\rm OCM}_{2\alpha} \leq -0.4$ MeV, 
than the $2\alpha$ RGM. If we readjust the potential parameter
to fit the $2\alpha$ binding energy, we would apparently obtain
an even worse result. It was realized a long time ago
that a simple choice of the direct potential $V_{\rm D}$ for
the effective interaction $V^{\rm eff}$ in OCM gives
a poor result. \cite{TA81}

\section{Summary}

The main purpose of this study is to find an optimum equation
for three-cluster systems interacting via pairwise two-cluster
RGM kernel. This is a necessary first step to apply the quark-model
baryon-baryon interactions to few-baryon systems.
We have found that the orthogonality condition to the pairwise
Pauli-forbidden states is a compulsory condition to assure a reasonable
result. The inherent energy dependence of the two-cluster exchange
RGM kernel should be treated self-consistently, if the eigen-values
of the exchange normalization kernel $K$ are non-negligible
in the allowed space.
The proposed three-cluster equation has a nice feature that
the equivalent three-cluster Faddeev equation is straightforwardly
formulated using the non-singular part of the full $T$-matrix
derived from the RGM kernel.
We have applied this equation to simple systems composed
of the three di-neutrons and three $\alpha$ clusters.
The equivalent Faddeev equation is also solved for the three
di-neutron system.
The Faddeev calculation for the $3\alpha$ system will be
reported in a separate paper. \cite{FU02}
For the ground state of the three $\alpha$ system,
the obtained binding energy is 1.5 MeV less bound,
in comparison with the full microscopic three-cluster RGM calculation. 
We think this satisfactory, since three-cluster RGM calculation
of the few-baryon systems using quark-model baryon-baryon interaction
is still beyond the scope of feasibility.
The application to the hypertriton using our quark-model
nucleon-nucleon and hyperon-nucleon interactions \cite{FU96b,FU01b}
is now under way.


\section*{Acknowledgements}

This work was supported by a  Grant-in-Aid for Scientific
Research from the Ministry of Education, Science, Sports and
Culture (No. 12640265).


\appendix


\begin{table}[ht]
\caption{The lowest eigen-values for $3d^\prime$ and $3\alpha$ systems,
obtained by diagonalization using [3] symmetric
translationally-invariant h.o. basis. 
The orthogonality condition by the projection
operator $P$ in \protect\eq{three3} is imposed.
$N$ stands for the maximum total h.o. quanta included in the calculation,
and $n_{\rm Max}$ the number of the basis states with the orbital
angular momentum $L=0$.
The three-cluster equation \protect\eq{three1} is used
for $E_{3d^\prime}$ and $E_{3\alpha}$, while the energy-independent
renormalized interaction \protect\eq{alpha2} and $v=\Lambda
(V_{\rm D}+V^{\rm CL}_{\rm D}) \Lambda$ are used in \protect\eq{three5}
for $E^{\rm RN}_{3\alpha}$ and $E^{\rm OCM}_{3\alpha}$, respectively.
}
\label{table}
\begin{center}
\renewcommand{\arraystretch}{1.1}
\setlength{\tabcolsep}{4mm}
\begin{tabular}{c|cc|ccccc}
\hline
 & \multicolumn{2}{c|}{$3d^\prime$} 
 & \multicolumn{5}{c}{$3\alpha$} \\
\cline{2-8}
$N$ & $n_{\rm Max}$ & $E_{3d^\prime}$
 & $n_{\rm Max}$ & $\varepsilon$ & $E_{3\alpha}$
 & $E^{\rm RN}_{3\alpha}$ & $E^{\rm OCM}_{3\alpha}$ \\
\hline
4  &  1 & 3.256  & $-$ & $-$ & $-$ & $-$ & $-$ \\
6  &  3 & 2.828  & $-$ & $-$ & $-$ & $-$ & $-$ \\ 
8  &  6 & 0.7373 &  1  & 25.120 & 12.634 &  12.634 & 23.570  \\
10 & 10 & 0.5585 &  3  & 18.023 &  3.615 &   4.575 & 11.422  \\
12 & 16 & 0.1169 &  7  & 14.857 & $-0.343$ & 0.874 &  5.322  \\
14 & 23 & 0.0523 &  12 & 13.046 & $-2.454$ & $-1.194$ &  1.827 \\
16 & 32 & $-0.0868$ &  19 & 11.920 & $-3.678$ & $-2.449$ & $-0.323$ \\
18 & 43 & $-0.1351$ &  28 & 11.182 & $-4.439$ & $-3.255$ & $-1.703$ \\ 
20 & 56 & $-0.1972$ &  39 & 10.682 & $-4.929$ & $-3.788$ & $-2.614$ \\
22 & 71 & $-0.2313$ &  52 & 10.339 & $-5.252$ & $-4.148$ & $-3.227$ \\
24 & 89 & $-0.2660$ &  68 & 10.099 & $-5.470$ & $-4.394$ & $-3.647$ \\
26 & 109 & $-0.2899$ & 86 &  9.931 & $-5.619$ & $-4.565$ & $-3.939$ \\
28 & 132 & $-0.3117$ & 107 & 9.812 & $-5.721$ & $-4.685$ & $-4.143$ \\
30 & 158 & $-0.3284$ & 131 & 9.727 & $-5.792$ & $-4.769$ & $-4.289$ \\
32 & 187 & $-0.3431$ & 158 & 9.666 & $-5.842$ & $-4.829$ & $-4.394$ \\
34 & 219 & $-0.3550$ & 188 & 9.623 & $-5.878$ & $-4.872$ & $-4.469$ \\
36 & 255 & $-0.3653$ & 222 & 9.591 & $-5.903$ & $-4.904$ & $-4.525$ \\
38 & 294 & $-0.3740$ & 259 & 9.568 & $-5.921$ & $-4.926$ & $-4.565$ \\
40 & 337 & $-0.3815$ & 300 & 9.551 & $-5.934$ & $-4.943$ & $-4.595$ \\
42 & 384 & $-0.3879$ & 345 & 9.539 & $-5.943$ & $-4.954$ & $-4.618$ \\
44 & 435 & $-0.3934$ & 394 & 9.530 & $-5.950$ & $-4.964$ & $-4.635$ \\
46 & 490 & $-0.3982$ & 447 & 9.524 & $-5.955$ & $-4.971$ & $-4.647$ \\
48 & 550 & $-0.4025$ & 505 & 9.519 & $-5.959$ & $-4.976$ & $-4.657$ \\
50 & 614 & $-0.4061$ & 567 & 9.515 & $-5.962$ & $-4.979$ & $-4.664$ \\
52 & 683 & $-0.4094$ & 634 & 9.512 & $-5.964$ & $-4.982$ & $-4.669$ \\
54 & 757 & $-0.4122$ & 706 & 9.510 & $-5.965$ & $-4.984$ & $-4.674$ \\
56 & 836 & $-0.4147$ & 783 & 9.508 & $-5.966$ & $-4.986$ & $-4.677$ \\
58 & 920 & $-0.4169$ & 865 & 9.507 & $-5.967$ & $-4.987$ & $-4.679$ \\
60 & 1010 & $-0.4189$ & 953 & 9.506 & $-5.968$ & $-4.988$ & $-4.681$ \\
\hline
\end{tabular}
\end{center}
\end{table}

\end{document}